# Residual Block-based Multi-Label Classification and Localization Network with Integral Regression for Vertebrae Labeling

Chunli Qin, Demin Yao, Han Zhuang, Hui Wang, Yonghong Shi, and Zhijian Song

*Abstract*—Accurate identification and localization of the vertebrae in CT scans is a critical and standard preprocessing step for clinical spinal diagnosis and treatment. Existing methods are mainly based on the integration of multiple neural networks, and most of them use the Gaussian heat map to locate the vertebrae's centroid. However, the process of obtaining the vertebrae's centroid coordinates using heat maps is non-differentiable, so it is impossible to train the network to label the vertebrae directly. Therefore, for end-to-end differential training of vertebra coordinates on CT scans, a robust and accurate automatic vertebral labeling algorithm is proposed in this study. Firstly, a novel residual-based multi-label classification and localization network is developed, which can capture multi-scale features, but also utilize the residual module and skip connection to fuse the multi-level features. Secondly, to solve the problem that the process of finding coordinates is non-differentiable and the spatial structure is not destructible, integral regression module is used in the localization network. It combines the advantages of heat map representation and direct regression coordinates to achieve end-to-end training, and can be compatible with any key point detection methods of medical image based on heat map. Finally, multi-label classification of vertebrae is carried out, which use bidirectional long short term memory (Bi-LSTM) to enhance the learning of long contextual information to improve the classification performance. The proposed method is evaluated on a challenging dataset and the results are significantly better than the state-of-the-art methods (mean localization error <3mm).

## I. Introduction

LOCALIZATION, labeling and segmentation of vertebrae and intervertebral discs are essential tasks in computer-assisted spinal surgery [1], and the accuracy and robustness of these tasks are critical for subsequent clinical tasks such as pathological diagnosis, surgical planning and post-operative evaluation [2][3][4]. Among them, vertebrae are the basic anatomical markers that, in addition to providing the important shape of the spine, serve as a reference structure for other organs, drawing more attention to the study of vertebrae. In spinal imaging, the identification and localization of the vertebrae is a pre-processing step for spinal analysis [5], segmentation [6][7], and registration [8]. It is also crucial for the clinical application of scoliosis, vertebral fracture [9], back pain [10] and other clinical tasks.

In medical imaging, CT scans have higher sensitivity and specificity in the visualization of bone structure, so they become an essential tool for diagnosing spinal anomalies [1]. By manually searching specific vertebrae in CT, such as the first cervical vertebra, i.e., C1, doctors can determine the type and location of subsequent vertebrae, but manual labeling is subjective and time-consuming [11]. When the field of view of CT is limited, for example, there is only the second to the tenth thoracic vertebrae in the image, i.e., T2-T10, the doctor cannot obtain the reference vertebrae, which makes it more difficult to identify and locate the vertebrae. Therefore, it is necessary to develop an automatic identification and localization method to improve the efficiency and reliability of vertebrae labeling.

However, there are great challenges in developing this method: as shown in Fig.1A, adjacent vertebrae have very similar appearance structures, which makes it challenging to discriminate different vertebrae. Generally, the field of view (FOV) of the spine CT image varies widely, and usually only a part of the entire spine is captured, so the image cannot provide complete contextual information. When there is metal implant in the CT image, it will increase the contrast around the bone boundary and cause image artifacts, as shown in Fig.1B. In addition, as shown in Fig.1C and Fig.1D, severe scoliosis and curvature of the spine increase the differences of spine morphology among the subjects, which increases the difficulty of vertebrae detection.

To solve the problems mentioned above, many methods have been proposed for successfully labeling vertebrae. Some early model-based methods relied on prior information of the vertebrae to identify the vertebrae by capturing the shape of the spine, but they lacked universality [12][13][14][15][16]. The later methods can be summarized into machine learning-based methods and deep learning-based methods according to the

C. Qin, D. Yao, Y. Shi and Z. Song are with the Digital Medical Research Center, School of Basic Medical Sciences, Fudan University, and also with the Shanghai Key Laboratory of Medical Imaging Computing and Computer Assisted Intervention, Shanghai, 200032,China, (e-mail: clqin@fudan.edu.cna; rh386@sina.com; yonghong.shi@fudan.edu.cn ; zjsong@fudan.edu.cn). Y. Shi and Z. Song are co-correspondents.

H. Zhuang and H. Wang are with the Digital Medical Research Center and Department of Anatomy, Histology and Embryology, School of Basic Medical Sciences, Fudan University, and also with the Shanghai Key Laboratory of Medical Imaging Computing and Computer Assisted Intervention, Shanghai, 200032, China (e-mail: zhuangh17@fudan.edu.cn; huiwang18@fudan.edu.cn).

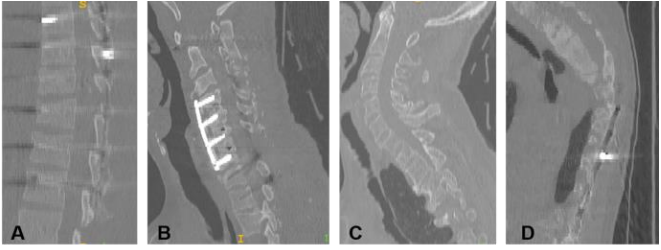

Fig. 1. Some challenges in vertebra identification and localization in CT scans. A. The adjacent vertebrae have the very similar appearance structure. B. There are the metal implants in the spine. C and D. There exist scoliosis and abnormal curvature of spine.

development stage.

*1) Machine learning-based methods*

For example, Glocker *et al.* [17] proposed a method based on regression forest and Markov process, but this method may not be suitable for pathological spine CT. In order to solve this problem, Glocker *et al.* [18] further proposed a robust forest-based classification algorithm, which avoided the modeling of shape and appearance and successfully labeled the vertebrae on normal and pathological spine CT. However, early methods used hand–crafted features, which are not sufficient to detect vertebrae when there are metal implants or spinal curvature in the image.

*2) Deep learning-based methods*

In order to automatically extract the robust features of vertebrae, many methods use a deep learning framework to solve the problem of vertebra detection. Chen *et al.* [19] proposed a joint convolutional neural network (J-CNN), which first uses a random forest to roughly locate candidates, and then combines paired information of adjacent vertebrae to eliminate false positives. This method is much better than previous methods, but they use 2D CNN to extract features, which may ignore the spatial information of CT scans, making the extracted features not very effective. In order to better identify and locate vertebrae, some scholars, inspired by human posture estimation methods, use Gaussian heat map-based methods to label vertebrae. Yang *et al.* [20][21] proposed an image-to-image 3D full convolutional network (DI2IN) with deep supervision to detect vertebrae, introduced messaging or convolutional LSTM to learn the long context information (spatial information along the direction of the spine from the current vertebrae) of vertebrae, and then used sparse regularization or shape dictionary to refine localization results. Similarly, Sekuboyina *et al*. [22] also used Gaussian heat map representation and proposed a 2D Butterfly Full Convolutional Network (Btrfly) based on the projection information of CT scans on sagittal and coronal planes, and encoded the spinal structure into Btrfly using an energy-based antagonistic automatic encoder. This method performs very well on the same dataset, however, they use the sagittal and coronal planes of 3D CT images, so some spatial information may be ignored.

The method based on heat map representation achieves very good performance, it generates Gaussian heat map for each vertebra in CT, and selects the position of the maximum probability value in the heat map as the predicted vertebra centroid. However, the operation of taking the coordinates of the maximum probability value is non-differentiable, so it is impossible to use network to carry out end-to-end training. Liao *et al.* [23] combined 3D classification and localization network to learn and share the features of the short spinal context information. Considering that the vertebrae in the spinal CT have a fixed spatial order, they subsequently combined the bi-directional LSTM (Bi-LSTM) to learn the long context information, achieving the best results. However, its localization network directly uses the fully connection layers to regress the coordinates, which may ignore the spatial information.

More and more deep learning-based methods are being used to identify and locate vertebrae. These methods mainly integrate multiple networks to improve the detection performance of vertebrae. In addition, most methods use Gaussian heat maps, which require utilizing the operation for maximum value to obtain the coordinates of the maximum response value on the heat map, but this method is non-differentiable and impossible to train the coordinates end-to-end using the network. Although the direct regression methods can carry out end-to-end differential training, it lacks spatial generalization ability. To address these issues, our contributions are as follows:

(1) We propose a novel end-to-end residual block-based multi-label classification and localization network, which can take into account both local and global information of vertebrae. Inspired by [24][25], we design the network to learn multi-scale features to label vertebrae, and use residual modules to fuse the multi-layer features to prevent the network gradient from disappearing.

(2) In the localization branch of the network, an integral regression module is added to convert the spatial heat map into coordinate points. It harmonizes the advantages of Gaussian heat map and direct regression coordinates, and has the ability of full differential training and spatial generalization. Different from the direct use of the heat map representation, our method carries out integral regression to the normalized heat map, and indirectly learns the heat map by optimizing the predicted coordinates of the model output, without directly applying the loss to the heat map, otherwise the predicted coordinates may not be accurate.

(3) In the classification branch of the network, multi-label learning and bidirectional LSTM are utilized to improve classification performance by learning the long context information far away from the current vertebra along the spine.

(4) The proposed method is trained and tested on the public challenging datasets, and the results show that our method is significantly better than the existing methods in performance.

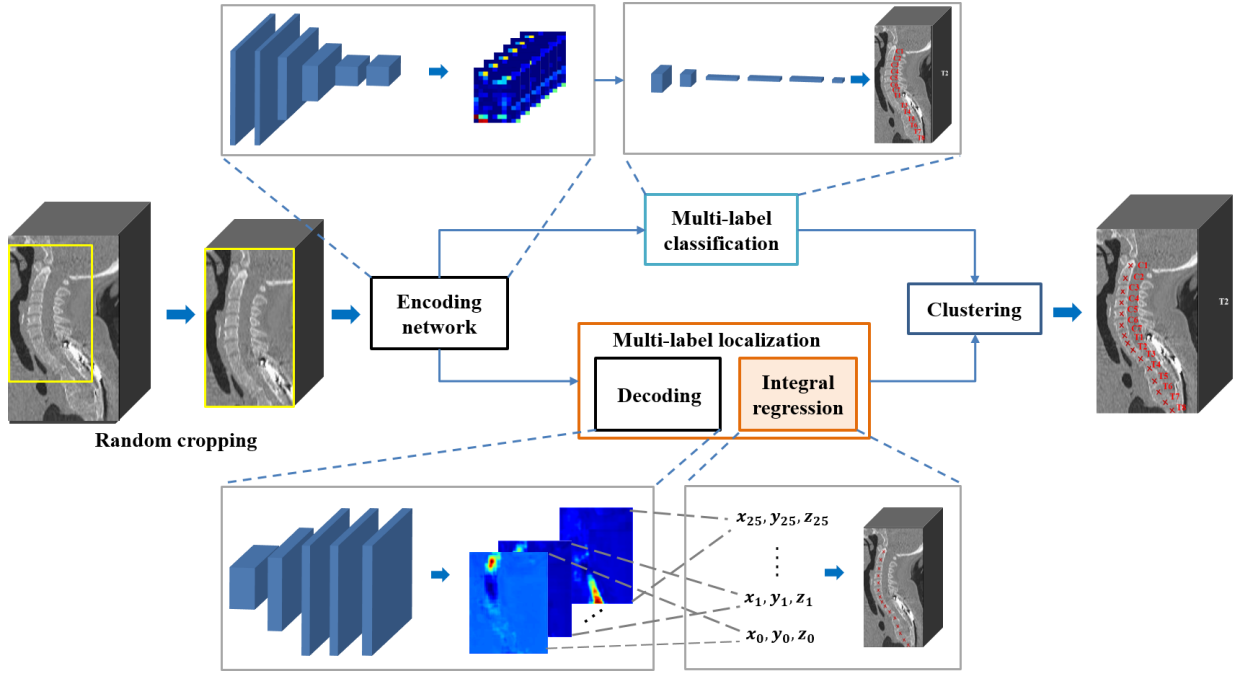

Fig. 2. The overall architecture of the method.

## II. METHODOLOGY

The overall architecture of our approach is shown in Fig.2. Any spine CT image is cropped into blocks of a certain size, and then input into the encoding network to extract multi-scale features, which are transferred to the classification and location branch, and finally the vertebra centroid coordinates classified as the same type are clustered to obtain the output results. We introduce the method in three stages. Firstly, the overall architecture of the novel residual block-based multi-label classification and localization network is introduced. Inspired by hourglass-net [24], we use multi-scale features to detect vertebra. Secondly, in the localization branch of the network, we added the integral regression module to improve the localization performance. Finally, we describe in detail the multi-label classification network, whose function is to improve the vertebra recognition rate by eliminating false positives.

### A. Residual block-based multi-label classification and localization network

Fig.3 details the residual block-based multi-label classification and localization network. The input of the network is a cropped 3D spine CT, and the output is a multi-channel vertebrae type and the predicted vertebrae centroid coordinates. The network consists of three parts：

(1) *The encoding network,* which is the classification and location sharing network being composed of a bottom-up sampling branch and four skip-out branches. The down-sampling branch has 6 residual modules and 3 max-pooling layers. Through this branch, the image is sampled from high resolution to low resolution to capture the features of the image at different scales, enabling different types of vertebrae to be distinguished, such as, cervical, thoracic, lumbar, and sacral.

(2) *The localization network branch,* which is an up-bottom process from low resolution to high resolution, with 3 residual modules, 3 up-sample operations, 2 convolutional layers and an integral regression module. This branch and the shared network form an hourglass network, which is similar to FCN [26] and U-net [27], and both have encoding and decoding process. However, the difference is that the proposed localization network is a symmetrical topological structure. On the skip-out branch, convolution operation is performed on the features of the original resolution, and the features of the same scale on the up-sampling branch are added element by element. After the network reaches the maximum scale, two continuous convolutional layers are applied to generate the pixel-level prediction, and the output Gaussian heat map is input into the integral regression module to calculate the predicted centroid coordinates of different types of vertebrae.

(3) *Multi-label classification branch.* This part has two convolutional layers (the number of kernel is 32, the filters are $5\times3\times3$ and $1\times1\times1$ respectively), one max-pooling layer, three bidirectional long short term memory network (Bi-LSTM) with 128 hidden states and a fully connection layer. The high level features extracted from the shared network are passed to the branch and the probability values of 26 vertebrae are finally output.

*Residual module*

A large number of residual [28] modules as shown in Fig. 4 are used in the network, which is composed of convolution branch and skip-out branch. The convolution branch is composed of a 1×1×1 convolution, a 3×3×3 convolution and a 1×1×1 convolution in series. Group Normalization [29] and Relu is used after each convolution to adjust the distribution of the output data in each layer of the network. The convolution branch of residual module extracts the features of higher level, while the skip-out branch retains the information of the original level. This module does not change the image scale, only changes the data depth (or the number of channels as shown in Fig.4), and can operate on images of any size. Moreover, it uses 1×1×1 convolution to reduce parameters, thus reducing the memory consumption.

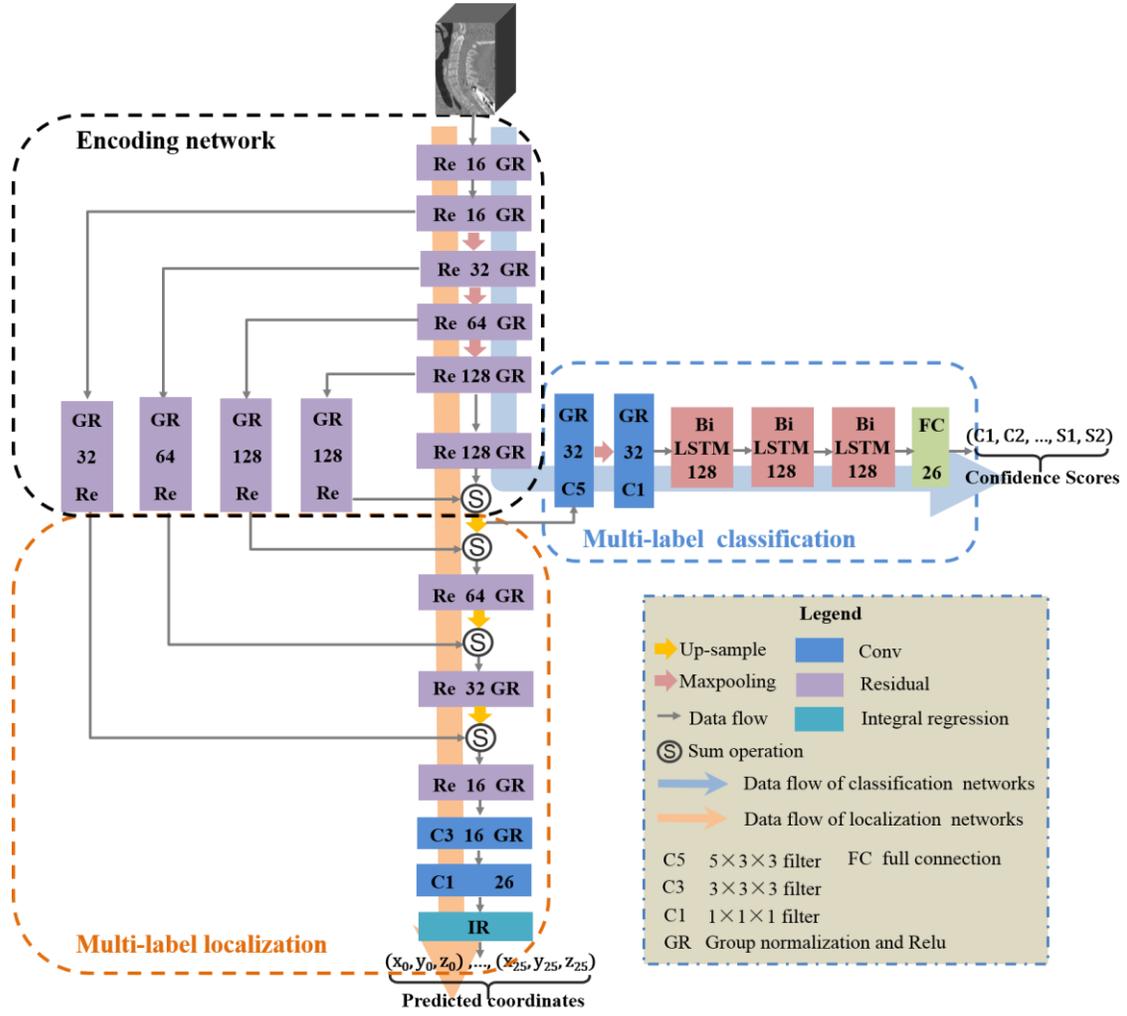

Fig. 3. The architecture of the residual block-based multi-label classification and localization network. The multi-label classification network outputs the confidence scores of the 26 vertebrae, and the multi-label localization network outputs the predicted centroid coordinates of the 26 vertebrae. The numbers on the residual module denote the number of channels for model output, the numbers on the convolution layer module denote the number of convolution kernels, the numbers on the Bi-LSTM denote the number of hidden states and the numbers on the fully connection layer denote the neuron number.

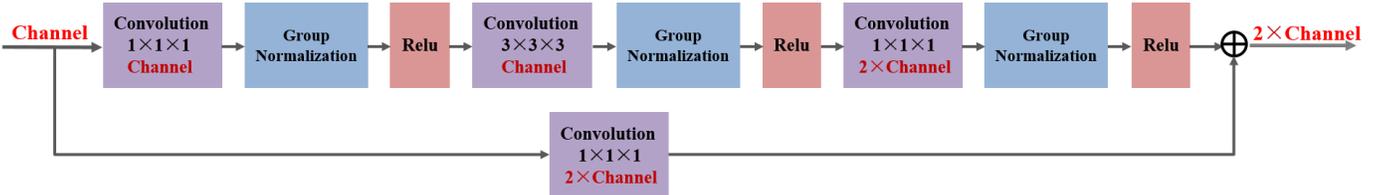

Fig. 4. Residual module structure. It is composed of three convolutional layers and jump layers. Group normalization and Relu are used after each convolution layer, and 1×1×1 convolution is used to reduce parameters.

*Loss function*

Generally speaking, the human body usually has 7 cervical vertebrae C1-C7, 12 thoracic vertebrae T1-T12, 5 lumbar vertebrae L1-L5, and 2 sacral vertebrae S1-S2 [1]. So for any spine CT image, let $v_n = (x_n, y_n, z_n), n = 0,1,2,…,25$ denote the coordinates of the centroid of the 26 vertebrae. Then the output of the classification network is 26 probability values, which respectively denote the probability that the CT image contains the corresponding vertebra. It can be defined as a 26-

dimensional vector, $\boldsymbol{u} = [u_0, \ldots, u_n, \ldots, u_{N-1}]$, $N = 26$, $u_n \in \{0,1\}$, $u_n$ indicates whether a corresponding vertebra exists in the CT scans. The zero vector, $u_n = 0$, means that there is no vertebra in CT. We use the crossing entropy loss as the loss function of classification. Due to the sparse labels, that is, the number of "1" is less than the number of "0", there is a classification imbalance problem, so a balance factor $B$ is introduced to enhance the learning of the vertebrae, here $B$ is 2 to 4.

$$L_{cls} = \sum_{u_n=1} -B \log(f_{cls}(v_n)) - \sum_{u_n=0} \log(1 - f_{cls}(v_n)). \quad (1)$$

The output of the localization network is 26 three-dimensional coordinates, which can be regarded as a regression problem. $Smooth_{L_1}$ which is commonly used in object detection [30], is used to represent the localization loss, and the non-spinal vertebra is ignored. The loss function is as follows:

$$L_{reg} = \lambda \sum_n \sum_i u_n smooth_{L_1}\left(t_i - f_{reg}^i(v_n)\right), \quad (2)$$

$$smooth_{L_1}(x) = \begin{cases} 0.5x^2, & \text{if } |x| < 1 \\ |x| - 0.5, & \text{otherwise} \end{cases}. \quad (3)$$

When the $n^{th}$ vertebra is present in CT scans, $u_n = 1$, $t_i$ denotes the $i^{th}$ dimensional coordinates of the $n^{th}$ vertebra, and $f_{reg}^i(v_n)$ denotes the predicted vertebra centroid coordinates. We set $\lambda = 0.4$ to approximately equal the number of categories and regressions.

The network trains classification and localization at the same time. The total loss function $Loss_{total}$ is defined as the combination of all output losses, as shown below:

$$Loss_{total} = L_{cls} + L_{reg}. \quad (4)$$

*B. Multi-label localization empowered by integral regression*

At present, the centroid coordinates of the vertebrae need to be determined in the vertebrae detection. Essentially, it's a numerical regression. Specifically, regression of vertebral coordinates can be divided into two methods: (1) Use the fully connection layers to directly regress coordinates. As shown in Fig.5(a), this method uses the fully connection layer to directly output the vertebral coordinates based on the features of the CNN output, which can quickly perform end-to-end differential training. However, this method will greatly lose the spatial generalization ability, and the weight obtained through the fully connection will largely depend on the distribution of training data, which can easily lead to overfitting. (2) Pixel value regression based on Gaussian heat map. As shown in Fig.5(b), this method outputs a Gaussian heat map of all vertebrae coordinates, that is, each channel predicts a heat map of the vertebra, and then obtains the coordinates by using the maximum operation to calculate the maximum value of each channel offline. This method disassociates the loss from the coordinate points and cannot perform end-to-end differential training on the coordinates. In addition, Euclidean distance is usually used for the regression loss of Gaussian heat maps, which may lead to the deviation in learning results.

In view of the advantages and disadvantages of the above two methods, as shown in Fig.5(c), we use the integral regression module [31][32] with spatial generalization ability and differential advantage. This module learns the Gaussian heat map indirectly by optimizing the output of the whole model to predict the loss of coordinates, and replaces the maximum operation in the heat map with the integral operation, which combines the advantages of heat map representation and direct coordinate regression to perform coordinates end-to-end training. The method can be compatible with any key point detection method based on heat map in the 2D or 3D medical image, and the module is free of training parameters and other calculations.

The specific implementation of integral regression is shown in Figure 6, whose input is normalized probability map $P_j$. The Softmax function is used to normalize the Gaussian heat map so that all elements are non-negative and the sum is 1:

$$P_j = \frac{e^{Y_{jk}}}{\sum_k e^{Y_{jk}}}, \quad (5)$$

where $Y_{jk}$ denotes the $k^{th}$ pixel value of the $j^{th}$ heat map, and $P_j$ denotes the probability map of the $j^{th}$ vertebrae centroid, and each pixel value in the probability map denotes the probability that this position is the vertebrae centroid. The integral regression module integrates the normalized probability map, that is, each pixel value in the probability map is weighted with its corresponding coordinate to obtain the predicted centroid coordinate $C_j$ of the $j^{th}$ vertebrae.

$$C_j = \sum_z^D \sum_y^H \sum_x^B W_{zyx} \hat{Y}_j(z, y, x), \quad (6)$$

where D, H and B denotes the depth, height and width of the CT scans respectively, W denotes the coordinate matrix, which represents the coordinates (x, y, z) of the pixel in the probability map, and the size is D×H×B×3.

*C. Multi-Label classification empowered by Bi-LSTM*

Liao *et al.* [23] also trained a classification and localization network at the same time, which focuses on the determination of each independent label, that is, assuming that each label exists independently. In other words, for each cropped image, only a single label could match it. Therefore, the input of the network is a single vertebra CT, so that the coordinate point closest to the image center is used as the label of the sample (vertebra type and centroid). Due to the large range of vertebra sizes, in reality, there is usually more than one center of vertebra
in the fixed-size image cropped, so the label is not accurate for the image. The image cropped by our method contains multiple vertebrae centroids. As shown in Fig. 7, the size of image blocks cropped along the axial, coronal and sagittal directions of spine is $104 \times 80 \times 80$, and each vertebra has corresponding sample labels, that is, the multi-label learning method [33] is used. Multi-label learning methods can be generally divided into two categories, namely "problem transformation" method and "algorithm adaptation" method. We use the "problem transformation" method, the main idea of which is to transform the multi-label learning problem into other relatively simple problems by processing the given training data set, and then solve the given problem. For the study, we regard the classification of multiple vertebrae as the classification of vertebrae one by one, train an independent classifier for each vertebrae classification label with all samples in the training data set.

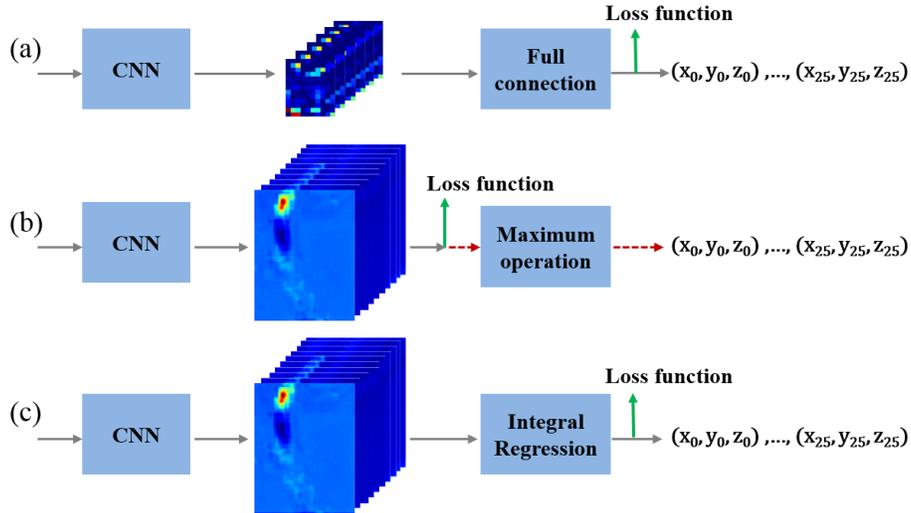

Fig. 5. Comparison of vertebrae centroid localization methods. (a) Using fully connection layers. (b) Using Gaussian heat map representation. (c) Using integral regression. layer, and $1\times1\times1$ convolution is used to reduce parameters.

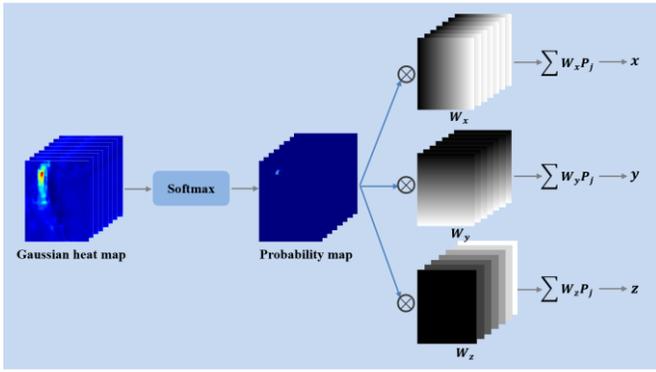

Fig. 6. Integral regression method. The Softmax function normalizes the Gaussian heat map to output a probability map. Each pixel value in the probability map is weighted with its corresponding coordinates to obtain (x, y, z), which denote the predicted centroid coordinate.

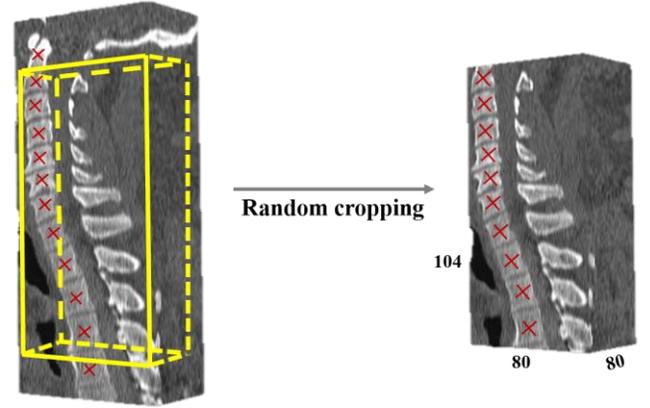

Fig. 7. Schematic illustration of random cropping of CT scans. The image size cropped along the axial, coronal and sagittal axes of the spine was $104 \times 80 \times 80$.

Although it is easy to decompose multi-label classification into multi-single label problems by training multiple single-label classification models, the correlation between vertebrae is ignored. The relationship between adjacent vertebrae can contribute to the classification of vertebrae, while the convolutional layer only learns short contextual information near the current vertebra. In order to learn the long context information far away from the current vertebra, as shown in Fig. 3, we used bi-directional LSTM [34] to learn the long sequence relationships near the current vertebra before using fully connection layers. In each time step of a bi-directional LSTM, the input is a one-dimensional feature vector extracted from the previous network. Here, we concatenated three bi-directional LSTMs, each with 128 hidden states.

## III. EXPERIMENTS

### A. Dataset and pre-processing

We evaluated the proposed approach on the 2014 MICCAI data set for vertebrae identification and localization. The data set included CT scans of 302 patients with different lesion types, among which 242 scans are for training and 60 for testing. Through screening, we excluded the images containing abnormal number of vertebrae, leaving 240 for training and 56 for testing. Most CT scans cover only a portion of the entire spine, and the problems such as abnormal curvature of the spine, metal implant, and limited field of view make it difficult to accurately identify and locate the vertebrae. The dataset provides the corresponding type and centroid coordinates for each vertebra.

In the data preprocessing stage, for each spine CT image, we limit the pixel value below the Hounsfield value of the air to -1000, and normalize the image with the mean value of 0 and the standard deviation of 1. The CT image was resampled to be isotropic with a resolution of 2mm×2mm×2mm, and the centroid coordinates of the vertebra were obtained at this resolution. In order to enhance the robustness of the network, we randomly cropped the original CT with a size of 104×80×80 along the axial, coronal and sagittal axes. The cropped image contained multiple continuous vertebrae and surrounding spatial information, whose size was limited by the network and memory. All experiments were conducted on two GTX Geforce 1080Ti GPUs using the Keras platform.

TABLE I
COMPARISON OF THE OVERALL LOCALIZATION PERFORMANCE OF THE VERTEBRAE. THE LOCALIZATION PERFORMANCE OF THE METHOD USING INTEGRAL REGRESSION WAS COMPARED WITH THAT USING HEAT MAPS (AND THEIR POST-PROCESSING METHODS) AND DIRECT REGRESSION, AND THE EVALUATION MEASURE USES MEAN AND STANDARD DEVIATION. (UNIT: MILLIMETER).

| Method | DI2IN[21] | DI2IN+MP[21] | DI2IN+ConvLSTM[20] | DI2IN+MP+Sparsity[21] | Btrfly Network[22] | 3D CNN LOC[23] | Integral Regression |
|---|---|---|---|---|---|---|---|
| Mean | 13.6. | 10.2 | 8.7 | 8.6 | 6.3 | 7.05 | 2.9 |
| Std | 37.5 | 13.9 | 8.5 | 7.8 | 4.0 | - | 3.1 |

TABLE II
COMPARISON OF CLASSIFICATION PERFORMANCE.

| Method | mAP(%) |
|---|---|
| Classification without Bi-LSTM | 85.6 |
| Classification with Bi-LSTM | 90.3 |

TABLE III
PERFORMANCE COMPARISON BETWEEN SINGLE LOCATION NETWORK AND PROPOSED MULTI-LABEL CLASSIFICATION AND LOCATION NETWORK..

| | Id. Rate (%) | Mean (mm) | Std (mm) |
|---|---|---|---|
| Localization network | 62.7 | 2.9 | 3.1 |
| Localization and Classification network | 89.0 | 2.3 | 2.8 |

*B. Train and test*

During the training phase, the Adam optimizer is used with an initial learning rate $\lambda = 0.01$ and the adaptive adjustment is used to reduce the learning rate with a coefficient of 0.4 and a minimum learning rate of $1 \times e^{-6}$ when the loss of the validation set stops decreasing. Considering memory limitations, we let batch size=2 and use group normalization (groups= 4) and weight decay $=1 \times e^{-4}$ to prevent overfitting. All network variables confirm convergence on the validation set. During the test phase, we conducted K-Means clustering algorithm on the coordinates of the same type of vertebrae output from the network, and the clustering results were used as the predicted centroid coordinates of the corresponding vertebrae types in the test CT scans.

*C. Evaluation metrics*

The mean average precision (mAP) was used to evaluate the performance of multi-label classification. First, the average precision (AP) of each category was calculated, and then the AP of all categories was averaged to measure the performance of the learned classifier in all categories. Two evaluation measures defined in [17] were also used to evaluate our overall network performance, namely recognition rate and localization error. If the estimated centroid closest to the label corresponds to the correct vertebra and the localization error is less than 20mm, the identification is considered correct. The localization error refers to the distance (the unit is millimeter.) of each predicted vertebra location from its label coordinates.

*D. Effect of integral regression*

In order to prove the effectiveness of integral regression module for vertebra localization, we compare the localization network with integral regression module with other methods based on heat map and direct regression coordinate. We removed the multi-label classification loss in the network, the parameters remain unchanged, and train the localization network. The results are shown in Table I.

As can be seen from Table I, the mean localization error of integral regression method is 2.9mm and the standard deviation is 3.1mm, which is obviously better than other methods. Similar to our localization network, DI2IN [21] used the full convolutional layers, but its network directly regressed the heat map pixel values and maximum operation was used to obtain the predicted coordinates. The mean error only reached 13.6mm, which was much worse than our result using integral regression. DI2IN+ConvLSTM [20] combined multiple networks to improve localization performance. After using a full convolutional network, they trained the ConvLSTM and Shape Basis networks to refine the localization results. DI2IN+MP [21] and DI2IN+MP+Sparsity [21] used message passing scheme and Sparsity regularization post-processing operations to suppress outliers after the network output heat map. Although the localization error is reduced, it is still more than twice that of the integral regression method which has no post-processing operation. Btrfly Network [22] uses a method similar to full convolutional networks based on heat maps. They also used independent network without any post-processing and the best mean localization error was 6.2mm. In the localization network, we used the integral regression module to implicitly carry out integral regression on the heat map to obtain the predicted vertebrae centroid, and conducted end-to-end differential training on the target coordinates without any post-processing operation. The mean error of 3D CNN LOC [23] is 7.05mm, which directly regress the predicted centroid coordinates of the vertebrae by using the fully connection layers. The reason why the localization effect was worse than the integral regression method was that the use of fully connection layers in the localization network damaged the network space generalizatio ability and easily lead to overfitting. From the comparison results, it can be seen that the use of integral regression module after the convolutional layer greatly optimizes the vertebrae localization effect, which has the advantages of spatial generalization ability and differential training.

TABLE IV
PERFORMANCE COMPARISON OF OUR METHOD WITH OTHER METHODS ON THE SAME DATASET.

| Method | Id. Rate (in %) | | | | | Mean(Std) (in mm) | | | | |
|---|---|---|---|---|---|---|---|---|---|---|
| | ALL | Cer. | Tho. | Lum. | Sac. | ALL | Cer. | Tho. | Lum. | Sac. |
| Chen[19] | 84.2 | 91.8 | 76.4 | 88.1 | - | 8.8(13.0) | 5.1(8.2) | 11.4(16.5) | 8.2(8.6) | - |
| Yang[20] | 85 | 92 | 81 | 83 | - | 8.6(7.8) | 5.6(**4.0**) | 9.2(7.9) | 11.0(10.8) | - |
| Liao[23] | 88.3 | **95.1** | 84.0 | 92.2 | - | 6.5(8.6) | 4.5(4.6) | 7.8(10.2) | 5.6(7.7) | - |
| Btrfly[22] | 86.7 | 89.4 | 83.1 | 92.6 | - | 6.3(**4.0**) | 6.1(5.4) | 6.9(**5.5**) | 5.7(6.6) | - |
| Btrfly$_{pe-w}$[22] | 87.7 | 89.2 | 85.8 | **92.9** | - | 6.4(4.2) | 5.8(5.4) | 7.2(5.7) | 5.6(6.2) | - |
| Btrfly$_{pe-eb}$[22] | 88.5 | 89.9 | 86.2 | 91.4 | - | 6.2(4.1) | 5.9(5.5) | 6.8(5.9) | 5.8(6.6) | - |
| Ours | **89.0** | 90.8 | **86.7** | 89.7 | 96.9 | **2.9**(5.8) | **2.2**(5.6) | **3.4**(6.5) | **2.9**(4.3) | 2.2(2.7) |

### E. Performance of multi-label classification

We used mAP to evaluate the performance of multi-label classification. Since the Bi-LSTM is used in the network to enhance the network learning of the long context information along the direction of the spine away from the current vertebra to improve the classification effect, as shown in Table II, the use of Bi-LSTM improved the performance of multi-label classification by nearly 5%.

### F. Results on classification and localization network

Using integral regression module in the localization network can achieve a good localization effect, but its ability to identify the type of vertebrae is not very good. To improve the recognition rate of vertebrae in the cropped CT scans, we constructed the novel multi-label classification and localization network to recognize and locate vertebrae simultaneously. The results are shown in Table III. The results showed that compared with the single localization network, the recognition rate (denoted as Id. Rate) of vertebrae was greatly improved after the addition of the classification network, because the false positivity of localization was eliminated by classification, and the localization effect of vertebrae was also improved.

Table IV shows the comparison results between our method and other latest methods on the same test data set, including the recognition rates, mean localization errors and standard deviation for all vertebrae and different types of vertebrae. As can be seen from Table IV, the mean localization error of the proposed method is far better than those of other methods, and the overall recognition rate of the vertebrae is the best. Liao *et al.* [23] used multi-task learning similar to our method, but they trained a single vertebra. To learn the correlation between vertebrae, they subsequently trained the Bi-LSTM network, compared with the previous method, the detection effect has been greatly improved, but the use of the fully connection layers impair the spatial generalization ability of the localization network and lose spatial information. Our method is an independent and self-contained multi-label network that simultaneously learns local and global context information for multiple vertebrae and performs end-to-end training. However, as can be seen from Table IV, compared with other types of vertebrae, thoracic vertebra has the worst recognition rate and localization error, because thoracic vertebra has the most types, and the appearance of other thoracic vertebrae is very similar except T1 and T12, which cannot be easily distinguished. Moreover, the thoracic vertebra has the most metal implants in the data set, making detection more difficult.

In the original CT of the test set, the localization error of the vertebrae increased, because the localization results of the test images are obtained by clustering the coordinates of all vertebrae classified into the same type in the cropped CT. When there are false positives in classification, that is, there are more classification errors, which have great influence on the results of localization. We will carry out further research to improve the performance of multi-label classification.

To further illustrate the effect of residual block-based multi-label classification and localization networks, some qualitative results are shown in Fig.8 and Fig.9. It can be seen from Figure 8 that our method can correctly identify and locate the cervical, thoracic, lumbar and sacral vertebrae, in which the blue crosses denote the ground truth and the red crosses denote our test results. In addition, our method can also correctly detect the vertebrae when the CT contains metal implants (Fig.8A), limited visual field (Fig.8B), and the vertebrae are similar in shape (Fig.8C.)

Fig.9 shows the visualization results with some problems (the blue crosses denote the ground truth, and the red crosses denote predicted vertebra locations). In Fig.9A, because the field of view of the image is highly restricted, the image contains metal implants, and the image boundary is blurred, this makes the classification method mistakenly classify C3 as background, resulting in the absence of C3 in the visualization results (green dotted box). Fig.9B shows that the localization results have a shift phenomenon (green dotted box), which is due to the high similarity of adjacent vertebrae. The abnormal curvature of the vertebrae in Fig.9C also leads to inaccurate localization results (green dotted box). Although there are vertebrae that failed to detect in some CT scans, for the entire CT, the number of failed vertebrae is small, and the overall

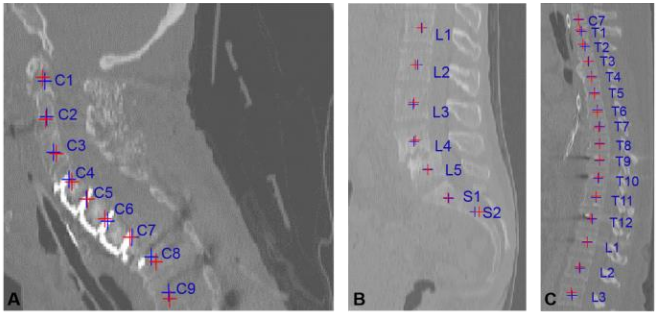

Fig. 8. Visualization results on different spine types using the proposed method. The blue rosses denote the ground truth and the red crosses denote our test results

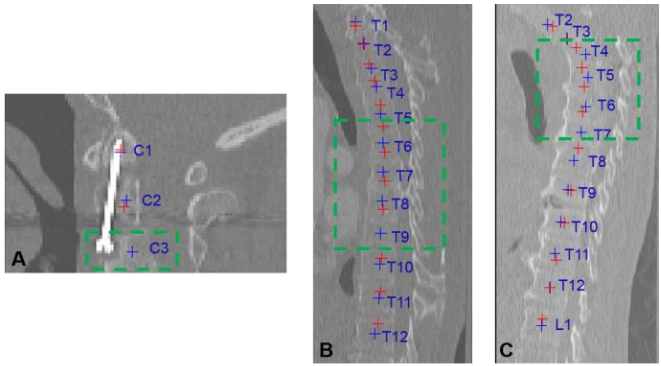

Fig. 9. Visual results with detection failure. The blue crosses denote the ground truth, and the red crosses denote the predicted results. The red dotted boxes denote the vertebra that failed to locate.

vertebrae detection results are acceptable.

## IV. CONCLUSION

In the study, we propose an effective method to identify and locate vertebrae in CT. We develop a novel residual block-based multi-label classification and localization network, which can take into account both local and global information of the vertebrae, learn features of different scales and fuses them. Each of the two network branches has its own advantages and shares information. On the localization branch, the integral regression module is used to learn the centroid coordinates of the vertebrae, which has the advantages of spatial generalization and end-to-end differential training, significantly reducing the localization error, and it can be compatible with any Gaussian heat map based medical image key point detection methods. In order to improve the recognition rate of the vertebrae, a multi-label classification network is trained while training the localization network, which can learn the short and long context information of the vertebrae at the same time. The proposed method is trained and evaluated on public challenging datasets, and the experimental results show the proposed method is significantly better than the state-of-the-art methods. In the future, we plan to improve the performance of multi-label classification networks on the basis of low localization errors and further improve the vertebrae recognition rate.

## ACKNOWLEDGMENT

This research was supported by grants from the National Key Research and Development Program of China (2017YFC0110701). This research was also supported by the National Natural Science Foundation of China (60972102).

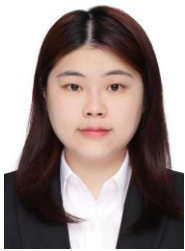

**Chunli Qin** received her Bachelor degree in Telecommunications engineering from Ocean University of China in 2017.She is currently a graduate student in School of Basic Medical Science of Fudan University. Her research interests are medical image processing and deep learning.

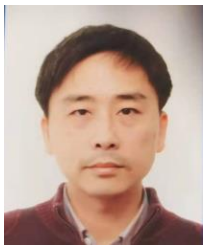

**Demin Yao** is currently working in Fudan University Digital Medicine Research Center and Shanghai Key Laboratory of medical image processing and computer-aided surgery. He has been engaged in medical image processing and related fields of scientific research, and has published over 10 scientific papers.

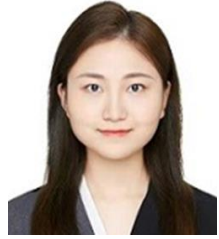

**Han Zhuang** received her Bachelor degree in medical imaging technology from Fujian Medical University of China in 2017. She is currently a graduate student in School of Basic Medical Science of Fudan University. Her research interest is medical informatics and medical image processing.

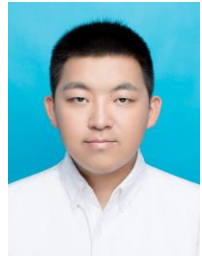

**Hui Wang** received his Bachelor degree in opto-electronics information science and engineering from Ocean University of China in 2018. He is currently a graduate student in School of Basic Medical Science of Fudan University. His research interests are medical image processing and deep learning.

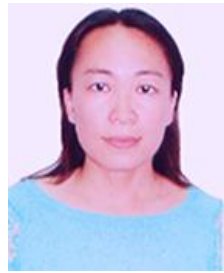

**Yonghong Shi**, since 2007, has been working in the school of basic medical sciences, Fudan University. She has been engaged in the research of medical image processing and computer-aided surgery. She has published more than 20 papers in IEEE TMI, IEEE TNNLs, MICCAI, Medical image analysis and other TOP journals and conferences.

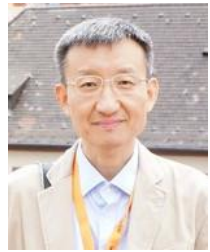

**Zhijian Song** received the B.S. degree from the Shandong University of Technology, Shandong, China, in 1982, the M.S. degree from the Jiangshu University of Technology, Jiangsu, China, in 1991, and the Ph.D. degree in biomedical engineering from Xi'an Jiaotong University, Xi'an, China, in 1994. He is currently a Professor in School of Basic Medical Science of Fudan University, Shanghai, China, where he is also the Director of the Digital Medical Research Center and the Shanghai Key Laboratory of Medical Image Computing and Computer Assisted Intervention (MICCAI). His research interests include medical image processing, image-guided intervention, and the application of virtual and augmented reality technologies in medicine.